\documentclass[twocolumn]{aastex62}

\newcommand{\Gaia}{{\it Gaia}}
\newcommand{\GALEX}{{\it GALEX}}

\newcommand{\Teff}{T_{\rm eff}}

\newcommand{\kms}{\ifmmode~{\rm km~s}^{-1}\else ~km~s$^{-1}~$\fi}

\newcommand{\bd}{BD+14$^\circ$3061}

\submitjournal{AJ}

\shorttitle{BD+14 3061}
\shortauthors{Bond}

\begin{document}          

\title{BD+14$^\circ$3061: A Luminous Yellow Post-AGB Star in the Galactic Halo}

\author{Howard E. Bond}  

\affil{Department of Astronomy \& Astrophysics, Pennsylvania State
University, University Park, PA 16802}
\affil{Space Telescope Science Institute, 3700 San Martin Drive,
Baltimore, MD 21218}
\affil{Visiting Astronomer, Cerro Tololo Inter-American Observatory, National Optical Astronomy Observatory, operated by the Association of Universities for Research in Astronomy under a cooperative agreement with the National Science Foundation.}

\begin{abstract}

I report the discovery that the 9th-magnitude Galactic-halo star \bd\ is a member of the rare class of luminous metal-poor ``yellow post-AGB'' stars. Its \Gaia\/ DR2 parallax implies an absolute magnitude of $M_V=-3.44\pm0.27$, and it is a very high-velocity star moving in a retrograde Galactic orbit. \bd\ is a field analog of the half-dozen yellow PAGB stars known in Galactic globular clusters, which have closely similar absolute magnitudes. These objects are the visually brightest members of old stellar populations; their { apparently} narrow luminosity function makes them potentially useful as Population~II standard candles. The spectral-energy distribution of \bd\ out to $22\,\mu$m shows no evidence for circumstellar dust. The star is a low-amplitude semi-regular pulsating variable, with typical periods of 30--32~days. A radial-velocity study suggests that it is a spectroscopic binary with a period of 429.6~days, making it similar to known binary yellow PAGB stars such as HD~46703 and BD+39$^\circ$4926. 

\null\vskip0.5in

\end{abstract}

\section{Yellow Post-AGB Stars}
 
The visually brightest objects in old populations are
post--asymptotic-giant-branch (PAGB) stars, evolving at nearly constant
luminosity to higher temperatures after leaving the AGB tip. As they pass
through spectral types early G, F, and late A, where the bolometric correction
is smallest, these ``yellow PAGB stars'' reach their brightest visual
luminosities.

I have argued \citep{Bond1997a, Bond1997b} that yellow PAGB stars in early-type galaxies and
halos of spirals are potentially excellent ``Population~II'' standard candles
for measuring extragalactic distances, because they are expected to have a very
narrow luminosity function. Moreover, they are easy to detect because they have
uniquely large Balmer discontinuities, arising from their very low surface
gravities. A group of colleagues at Pennsylvania State University is undertaking
a detailed study of these objects as standard candles, to be reported in a
forthcoming series of papers.

To calibrate the photometric zero-point, we primarily use yellow PAGB stars in
Galactic globular clusters (GCs) with well-established distances. However,
because of their short evolutionary timescales, these stars are extremely rare.
Only about a half-dozen are known in the Galactic GC system (e.g., \citealt{Bond2016}, hereafter B16, and references therein; Bond et al.\ 2020 in preparation). 

A few bright field analogs of these Population~II\footnote{ I am using the term ``Population~II'' to mean objects that appear to belong to an old stellar population, based on kinematics and/or a spatial location in the Galactic halo (or of course in a globular cluster). A low iron content is not, by itself, enough to place a PAGB star in an old population, because of the photospheric depletion processes described below.}  yellow PAGB stars are known.
Two prototypes are HD~46703 and BD+39$^\circ$4926 (see B16 and references
therein). The field stars are also potential contributors to the zero-point
determination, in cases where they are near enough for measurements of their
trigonometric parallaxes with high relative precision.

In this paper I report a bright yellow PAGB star in the field of the Galactic
halo. My discovery and initial observations were actually made many years ago,
but never published.\footnote{I did mention the discovery at a conference \citep{Bond1991}, and on this basis the star was included in a catalog of Population~II A-F
supergiants compiled by \citet{Bartkevicius1992}.} As noted above, we are now
studying this class of objects in detail, so it is appropriate to publish my
discovery now. I will present data I have accumulated on the star, along with
photometric data from all-sky surveys and precise astrometry from the {\it
Gaia\/} mission.

\section{BD+14$^\circ$3061: A New Yellow PAGB Star}

In the 1970s I carried out a survey for metal-deficient stars in the general
field, using objective-prism plates obtained with the Curtis Schmidt telescope
at Cerro Tololo Inter-American Observatory (CTIO)\null. I followed up by
obtaining photometry of the candidate stars in the Str\"omgren {\it uvby\/}
system. I presented details of the objective-prism survey,  a list
of extremely metal-poor red giants that were discovered, and the {\it uvby\/}
photometry, in an {\it Astrophysical Journal Supplements\/} paper (\citealt{Bond1980},
hereafter B80). 

In the course of examining a Curtis Schmidt plate I had obtained in 1977, I
classified the previously unstudied 9th-magnitude star BD+14$^\circ$3061 (hereafter shortened to
BD+14) as having an extremely weak-lined F-type spectrum. I was not able to
obtain Str\"omgren photometry until 1981. That measurement revealed that BD+14 is a new member of the
rare class of extremely low-gravity yellow PAGB stars. Table~\ref{table:basicdata} presents basic
data on the star, and results from my observations. The astrometric measurements
are from the \Gaia\/ Data Release~2 (DR2; \citealt{Gaia2018}).\footnote{\url{http://vizier.cfa.harvard.edu/viz-bin/VizieR-3?-source=I/345/gaia2}}

\begin{deluxetable*}{lcc}
\tablewidth{0 pt}
\tablecaption{Basic Data for BD+14$^\circ$3061
\label{table:basicdata}
}
%  \\
% \quad \bf NOTE: Preliminary version! Still being edited!
% \label{table:runs}}
\tablehead{
\colhead{Parameter} & 
\colhead{Value} &
\colhead{Source\tablenotemark{a}} 
}
\startdata
R.A. (J2000)     & 16:29:48.591  & (1) \\
Dec. (J2000)     & +14:15:43.15  & (1) \\
Parallax         & $0.251\pm0.032$ mas & (1) \\
R.A. proper motion & $-6.599\pm0.038\rm\,mas\,yr^{-1}$ & (1) \\
Dec. proper motion & $-17.148\pm0.035\rm\,mas\,yr^{-1}$ & (1) \\
Radial velocity & $-65.0\pm1.0\,\kms$ & (1) \\
Space velocity, $(U,V,W)$ & $(-120, -290, -51)\,\kms$ & (2) \\
$V$   & $9.490\pm0.012$ & (3) \\
$b-y$ & $0.341\pm0.007$ & (3) \\
$m_1$ & $0.044\pm0.013$ & (3) \\
$c_1$ & $1.159\pm0.018$ & (3) \\
Reddening, $E(B-V)$ & 0.053 & (4) \\
Absolute magnitude, $M_V$ & $-3.44\pm0.27$ & (5) \\
\enddata
\tablenotetext{a}{Sources: (1) \Gaia\/ DR2; (2) Space-velocity components
relative to the Sun ($U$ toward Galactic center, $V$ in direction of Galactic
rotation, $W$ toward Galactic north), calculated from \Gaia\/ astrometry and
radial velocity; (3) This paper; (4) {\tt Stilism} online tool (see text); (5) This paper, calculated from
data in this table (see text).}  
\end{deluxetable*}

\section{Str\"omgren Photometry}

\begin{figure}[ht]
\centering
\includegraphics[width=0.47\textwidth]{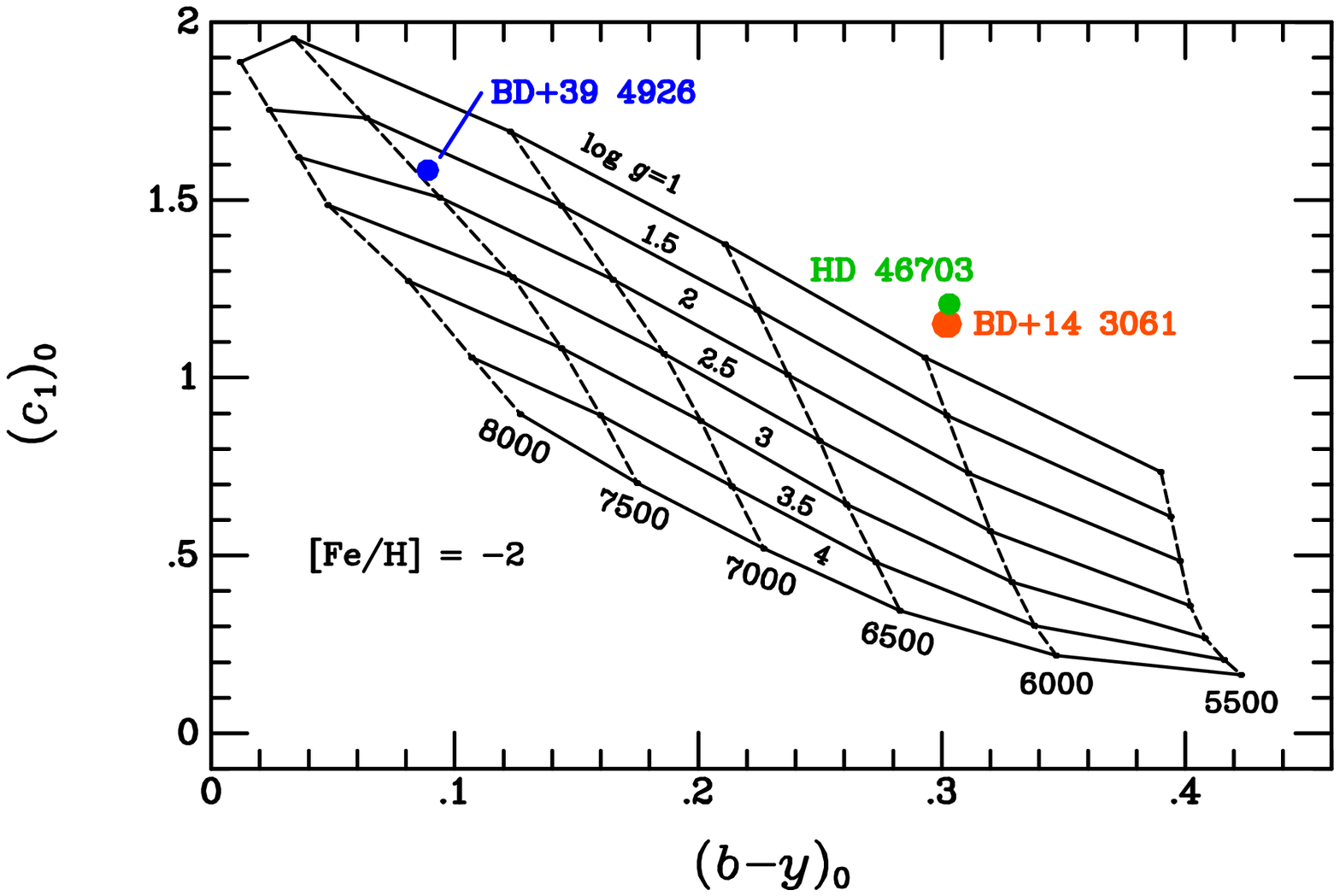}
\vskip0.2in
\includegraphics[width=0.47\textwidth]{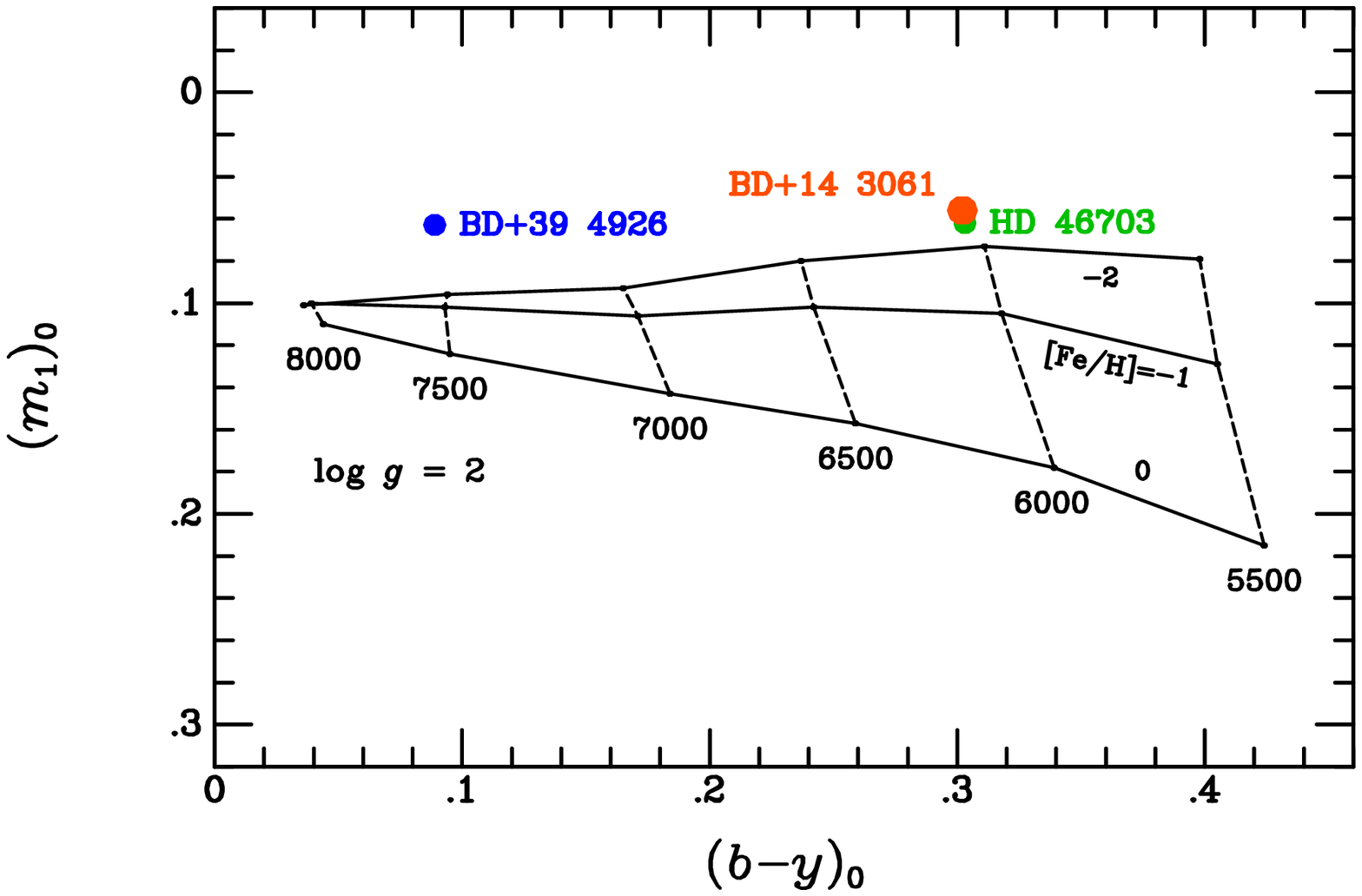}
\caption{
{\it Top panel:} Reddening-corrected gravity-sensitive color difference $(c_1)_0$ vs.\ corrected $(b-y)_0$ color
index for \bd\ (red) and two prototypical field yellow post-AGB stars: BD+39$^\circ$4926 (blue) and HD~46703 (green). Overlain is a grid of theoretical colors from \citet{Relyea1978} for a metallicity of $\rm[Fe/H]=-2$. 
{\it Bottom panel:} Reddening-corrected metallicity-sensitive color difference $(m_1)_0$ vs.\ $(b-y)_0$ for the three stars. Overlain is a grid of theoretical colors from Relyea \& Kurucz for a gravity of $\log g=2$.
\label{fig:uvby}
}
\end{figure}

I obtained photoelectric {\it uvby\/} photometry of BD+14 on 1981 April 11,
using the 0.9-m telescope at CTIO\null. I reduced the data to the standard
Str\"omgren system as described in B80, and they are presented in rows 6 through
9 in Table~\ref{table:basicdata}. The two panels in Figure~\ref{fig:uvby} plot the Str\"omgren $c_1$ and $m_1$ indices versus $b-y$ color for BD+14. Also plotted for comparison are data for
the prototypical stars HD~46703 (from \citealt{Luck1984}, hereafter LB84; the star is a
low-amplitude variable and the mean of their measurements is plotted) and
BD+39$^\circ$4926 \citep{Philip1973}. I corrected the photometry for all
three stars for interstellar reddening, using the {\tt Stilism} online
tool\footnote{\url{https://stilism.obspm.fr/}} \citep{Capitanio2017}, which estimates
reddening, $E(B-V)$, at any given Galactic position and distance: $E(B-V)=0.053$, 0.072, and 0.112 for BD+14, HD~46703,
and BD+39$^\circ$4926, respectively. I used the \citet{Crawford1975} relations between
reddening in the Str\"omgren and $BV$ systems to make the reddening corrections.

The top panel in Figure~\ref{fig:uvby} plots the gravity-sensitive $c_1$ index versus the
temperature index $b-y$. All three stars have very high values of $c_1$,
indicating extremely large Balmer discontinuities and low surface gravities. For
an approximate calibration, I have overlain the grid of theoretical colors from
\citet[][hereafter RK78]{Relyea1978}, using a metallicity of $\rm[Fe/H]=-2$.
The lowest gravity in the RK78 tables at this metallicity is $\log g=2$, so I
linearly extrapolated their values to $\log g=1.5$ and 1. The figure shows that
BD+14 has an extremely low gravity, below $\log g=1$, but the actual value is
uncertain because of the extrapolation. The colors of BD+14 are nearly identical
to those of HD~46703, for which LB84 used high-resolution spectra
to determine $\Teff=6000$~K and $\log g=0.4$; subsequently, \citet[][hereafter H08]{Hrivnak2008}
found 6250~K and 1.0 from an independent analysis. These values are in good
accord with the position of HD~46703 in the top panel of Figure~\ref{fig:uvby}. BD+14 thus
likely has very similar atmospheric parameters. For BD+39$^\circ$4926, a
high-dispersion analysis by \citet{Rao2012} gave $(\Teff, \log g)$ = (7750~K, 1.0),
in fairly reasonable agreement with Figure~\ref{fig:uvby}.

The bottom panel in Figure~\ref{fig:uvby} plots the metal index $m_1$ versus $b-y$ for BD+14
and the two prototype stars. All three stars have low $m_1$ values, indicative
of very weak metallic lines. In this panel I plot a grid of theoretical colors
from RK78 for metal contents of $\rm[Fe/H]=0, -1$, and $-2$. These colors are
for $\log g=2$, the lowest gravity in the RK78 tables, so they are only a rough
approximation for these very low-gravity stars. Nevertheless, it is clear that
all three stars have very low metal contents. BD+14 is again nearly identical
with HD~46703. LB84 found $\rm[Fe/H]=-1.6$ for this star. For
BD+39$^\circ$4926, the Rao et al.\ study gave $\rm[Fe/H]=-2.37$. 

The photospheres of HD~46703, BD+39$^\circ$4926, and several other PAGB stars show depletions of refractory chemical elements with high condensation temperatures (e.g., H08; \citealt{Oomen2018}, hereafter O18; \citealt{Oomen2019}; and references therein). This indicates that the refractory elements have been selectively removed from the gas through condensation onto grains in a cool region, with the gas then having been re-accreted onto the photosphere.\footnote{To my knowledge, I was the first to suggest this explanation for the photospheric depletions: \citet{Bond1991}.} A high-dispersion abundance analysis of BD+14 would be of considerable interest, to investigate whether it shows a similar effect.

\section{Space Motion, Absolute Magnitude, and Position in Color-Magnitude
Diagram}

Precise astrometry of BD+14 is available from \Gaia\/ DR2, as given in Table~\ref{table:basicdata}.
The measured parallax of $0.251\pm0.032$~mas, corrected by adding 0.029~mas
\citep{Lindegren2018}, implies a distance of $3.57^{+0.46}_{-0.37}$~kpc. Using
the DR2 proper motion and radial velocity (RV), from rows 4, 5, and 6 of
Table~\ref{table:basicdata}, I find a total space motion relative to the Sun of $318\,\kms$, making
BD+14 a star of very high velocity. The $U,V,W$ components of the star's space
velocity relative to the Sun are given in row 7 of Table~\ref{table:basicdata}; these show that
BD+14 is moving in a retrograde Galactic orbit.\footnote{ Since there is some evidence, presented below, that BD+14 is a long-period spectroscopic binary, it is possible that binary motion has affected the \Gaia\/ astrometry, although to some extent the perturbation may have averaged out over the duration of the \Gaia\/ observations. The DR3 data release will provide further information on this concern.}  

% ...remarkably close to Acheron stream in position \& distance.... but not
% precisely in it, after I got Sumner to calculate the lambda,eta coords

\begin{figure}[ht]
\centering
\includegraphics[width=0.48\textwidth]{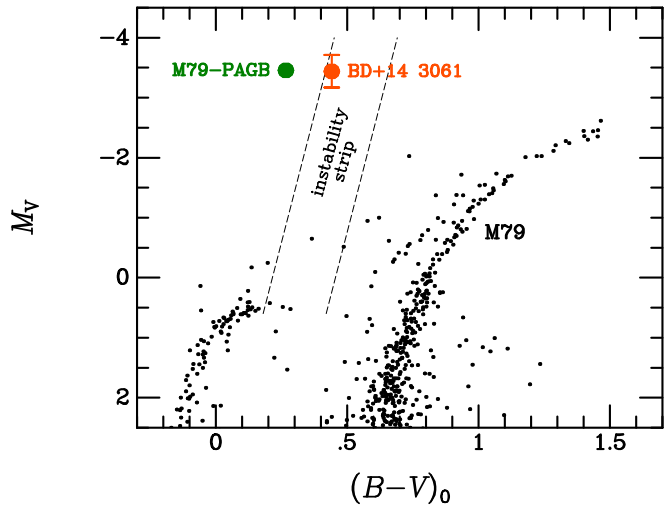}
\caption{
Dereddened color--absolute-magnitude diagram, $M_V$ versus $(B-V)_0$, for the yellow PAGB star \bd\ (red) and the PAGB star in the globular cluster M79 (blue). Also plotted is the color-magnitude diagram for M79 from \citet{Bond2016} (black dots). The location of the pulsational instability strip is marked. The M79 star is, by far, the visually brightest member of its cluster, and \bd\ is a nearly identical field analog, although sightly cooler and lying within the instability strip.
\label{fig:hrd}
}
\end{figure}

My measured $V$ magnitude for BD+14, adjusted for the distance and reddening in
Table~\ref{table:basicdata}, yields a visual absolute magnitude of $M_V=-3.44\pm0.27$. Figure~\ref{fig:hrd}
plots the position of BD+14 in the color versus visual absolute magnitude
diagram [with $b-y$ converted to $B-V$ using the approximate relation
$b-y\simeq0.69\,(B-V)$ from B80]. To place the star in context, I also plot the
color-magnitude diagram for the GC M79, taken from B16. As reported by B16, M79
contains a yellow PAGB star, and its position is also plotted in Figure~\ref{fig:hrd}. The
figure shows that the M79 PAGB star, at $M_V=-3.46$ (B16), has a nearly
identical absolute magnitude to that of BD+14. Both stars lie some four
magnitudes above the horizontal branch. However, BD+14 is cooler than the M79
PAGB star. The nominal location of the pulsational instability strip is indicated in the figure, and BD+14 appears to lie just within the blue edge of the strip. As noted below, it in fact is a low-amplitude variable.

The concordant visual absolute magnitudes found for BD+14 and the M79 star also agree very well with those of several more stars of this type known in GCs (\citealt{Alves2001}; B16, their Table~2; Bond et al.\ in preparation). All of them lie in the very narrow range $-3.10 \ge M_V \ge -3.46$.\footnote{However, in both GCs and the field, there are rare post--horizontal-branch stars, and post--early-AGB stars, which lie below the bright PAGB stars discussed here, and above the horizontal branch. (See, for example, \citealt{Harris1983}.) These include the Type~II Cepheids and RV~Tauri variables, as well as non-variables outside the instability strip. These objects will be explored in detail in a forthcoming paper (Davis et al., in preparation). The \Gaia\/ parallax of HD~46703 suggests it may belong to this class of ``above-horizontal-branch'' stars, but the relative uncertainty of its parallax is high, and thus its absolute magnitude is not well constrained. There is, in any case, an apparent sharp {\it upper\/} limit to the luminosities of the yellow PAGB stars in Population~II systems.} Thus the properties of BD+14 appear to strengthen the case that yellow PAGB stars are good candidates for Population~II standard candles.

\vskip0.2in

\section{Spectral-Energy Distribution}

Population~I PAGB stars are typically associated with substantial amounts of
circumstellar (CS) dust (e.g., \citealt{vanWinckel2003}; \citealt{deRuyter2006}, hereafter deR06, and
references therein). Dust is less commonly associated with Population~II yellow
PAGB stars like the ones discussed in this paper.  HD~46703 has only a weak
mid-infrared excess (deR06), with an IR luminosity of about 1\%
that of the photosphere. No IR excess has been detected for BD+39$^\circ$4926 (deR06). 

% The PAGB star in M79, for example, shows no evidence for circumstellar dust, as
% demonstrated by B16.

To test for the presence of CS dust around BD+14, I constructed its
spectral-energy distribution (SED) from the following data: (1)~{\it GALEX\/} \citep{Morrissey2007} far- and near-UV fluxes\footnote{\url{https://galex.stsci.edu/GalexView/}}; (2)~magnitudes\footnote{\url{https://www.aavso.org/apass}} in the Sloan Digital Sky Survey bandpasses ({\it ugriz\/}) from the AAVSO Photometric All-Sky Survey (APASS); (3)~my $V$ magnitude (Table~\ref{table:basicdata}); (4)~magnitudes\footnote{\url{https://irsa.ipac.caltech.edu/cgi-bin/Gator/nph-scan?submit=Select&projshort=2MASS}} in the 2MASS \citep{Skrutskie2006} near-IR bands ($J,H,K_s$); and (5)~magnitudes\footnote{\url{https://irsa.ipac.caltech.edu/cgi-bin/Gator/nph-scan?submit=Select&projshort=WISE}} in the four {\it WISE\/} \citep{Wright2010} mid-IR bands (3.4 to $22\,\mu$m). I corrected all of the measurements for reddening of $E(B-V)=0.053$, using the formulas of \citet{Cardelli1989}, with a total-to-selective extinction ratio of $R_V=3.1$.

\begin{figure}[ht]
\centering
\includegraphics[width=0.48\textwidth]{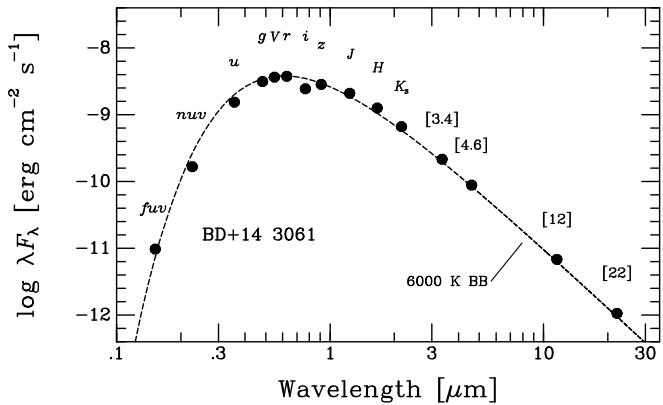}
\caption{
The spectral-energy distribution for \bd, corrected for interstellar extinction. The fluxes are taken from {\it GALEX\/} ({\it fuv\/} and {\it nuv}), APASS ({\it ugriz}), my $V$ magnitude, 2MASS ($JHK_s$), and {\it WISE\/} (3.4, 4.6, 12, and $22\,\mu$m). Also plotted is a blackbody with a temperature of 6000~K, normalized to the flux of \bd\ at the $V$ band. There is no evidence for significant circumstellar dust.
\label{fig:sed}
}
\end{figure}

The resulting SED is plotted in Figure~\ref{fig:sed}. For comparison I plot the flux for a
blackbody with a temperature of 6000~K, based on the photometry as discussed in
\S3. This blackbody provides an excellent fit to the SED\null. No IR excess is
seen out to $22\,\mu$m. The SED is very similar to that of the M79 PAGB star
(B16, their Figure~6), which likewise shows no evidence for CS dust. As
discussed by B16, dust formation may be difficult in low-metallicity stars, and
also the relatively long post-AGB lifetimes of these low-mass remnants is likely
to be long enough to allow any circumstellar dust to have dissipated. As also
noted by B16, the absence of CS material indicates that stars like BD+14 and the
M79 PAGB star are unlikely to produce ionized planetary nebulae during their
later high-temperature evolutionary phases. The lack of extinction by CS dust is also favorable for the utility of these stars as standard candles.

\section{Photometric Variability}

As noted above and shown in Figure~\ref{fig:hrd}, BD+14 appears to lie within the
pulsational instability strip, near its blue edge, in the color-magnitude
diagram. The similar and prototypical star HD~46703 is a known pulsating variable star (e.g.,
\citealt{Bond1984}; \citealt{Hrivnak2008}, hereafter H08); its photometric behavior is complex, with
a varying peak-to-peak amplitude of $\sim$0.1--0.38~mag, and a typical period of
about 29 to 31~days (H08).

\begin{figure}[ht]
\centering
\includegraphics[width=0.47\textwidth]{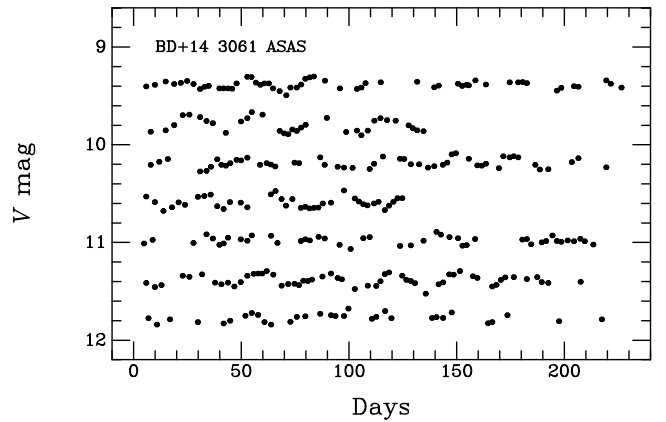}
\caption{
ASAS $V$-band light curves of \bd, for seven seasons from 2003 (top) to 2009 (bottom). The magnitudes on the $y$-axis apply to the 2003 season, and the light curves for the subsequent seasons have been shifted downwards successively by 0.4~mag. Pulsations with a period of $\sim$30--32~days are present, but with a variable amplitude, and sometimes nearly disappearing. 
\label{fig:lightcurve}
}
\end{figure}

To investigate whether BD+14 has similar light variations, I downloaded
photometric data obtained by the All Sky Automated Survey (ASAS; \citealt{Pojmanski1997}).\footnote{\url{http://www.astrouw.edu.pl/cgi-asas/asas_variable/162949+1415.7,asas3,0,0,500,0,0}} These observations are in the $V$ band, and cover annual seasons
from 2003 February to 2009 September. The ASAS light curves from these seven
seasons are plotted in Figure~\ref{fig:lightcurve}. 

As in the case of HD~46703, the star shows
complex variations, with a peak-to-peak amplitude that can reach $\sim$0.2~mag,
but which sometimes nearly disappear. This suggests an interaction between several
individual pulsation modes. A periodogram analysis indicates typical periods of
about 30--32~days. BD+14 is thus remarkably similar in its pulsation properties
to HD~46703. Both stars should perhaps be classified as members of the ``UU~Herculis'' group of variable stars (see \citealt{Sasselov1984}; \citealt{Zsoldos1993}), but they lack the strong IR excesses seen in most stars of the UU~Her class.

The mean magnitude for all of the ASAS observations is $V=9.39$, with which my
single measurement (Table~\ref{table:basicdata}) is in reasonable agreement. 

\section{Radial Velocities}

Both of the prototypical field old-population yellow PAGB stars mentioned in
this paper are long-period single-lined spectroscopic binaries. For HD~46703,
the orbital period is 597.4~days (\citealt{Waelkens1993}; H08; O18), and for BD+39 it is 871.7~days (\citealt{Kodaira1970}; O18).  The peak-to-peak RV amplitudes for these two stars are 33.7 and $30.6\,\kms$, respectively. There is a substantial literature (e.g., O18 and references therein) that associates membership in a wide binary with the selective depletion of elements in PAGB stars' atmospheres described above.

To check whether BD+14 is also a spectroscopic binary, I measured its RV
on 32 nights over the interval from 2003 February~21 to 2011 September~11.
Queue-scheduled spectrograms were obtained by Chilean service observers with 
the SMARTS\footnote{SMARTS is the Small \& Moderate Aperture Research Telescope System;  \url{http://www.astro.yale.edu/smarts}.} 1.5-m telescope at CTIO, using its RC-focus
spectrograph equipped with a CCD camera.

Two different grating setups were employed: (1)~setting ``56/II,'' covering
4017--4938~\AA, with a spectral resolution of 2.2~\AA, and (2)~setting
``47/IIb,'' covering 4070--4744~\AA, with a spectral resolution of 1.6~\AA\null.
Exposure times each night were generally $3\times75$~s or $3\times90$~s, and
short exposures of a HeAr calibration lamp were taken before and after each set
of stellar observations.

The CCD images were reduced as described by B16 for their observations of the
M79 PAGB star, which used the same telescope and spectrograph. The processing,
spectrum extraction, and wavelength calibration were all done using standard
IRAF\footnote{IRAF was distributed by the National Optical Astronomy Observatory, which is operated by the Association of Universities for Research in Astronomy (AURA) under a cooperative agreement with the National Science Foundation.} routines.

\begin{figure}[ht]
\centering
\includegraphics[width=0.47\textwidth]{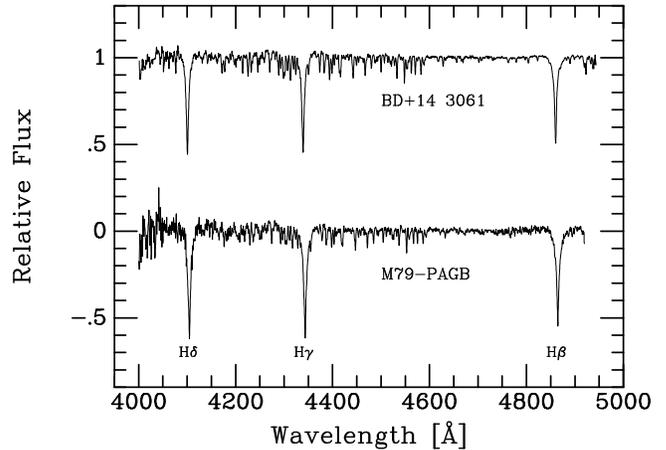}
\caption{
{\it Top panel:} Combined spectrum of \bd\ from 32 individual observations with the SMARTS 1.5-m spectrograph, normalized to a flat continuum.
{\it Bottom panel:} Combined spectrum from the same telescope for the PAGB star in the globular cluster M79. Three Balmer absorption lines are marked. All other lines are weak, consistent with very low metallicity. The two spectra are nearly identical.
\label{fig:meanspec}
}
\end{figure}

In Figure~\ref{fig:meanspec}, I show a spectrum created by combining all of the data for BD+14,
and normalizing to a flat continuum. For comparison, I also plot the combined
spectrum of the M79 PAGB star, taken from B16. The two spectra are nearly
identical, showing sharp, strong absorption lines of the Balmer series. All of
the metallic lines are quite weak in both stars. A high-dispersion analysis of
the M79 star yielded $\rm[Fe/H]=-2.0$ \citep{Sahin2009}, and the metallicity of BD+14
is likely to be similar.

\begin{figure}[ht]
\centering
\includegraphics[width=0.46\textwidth]{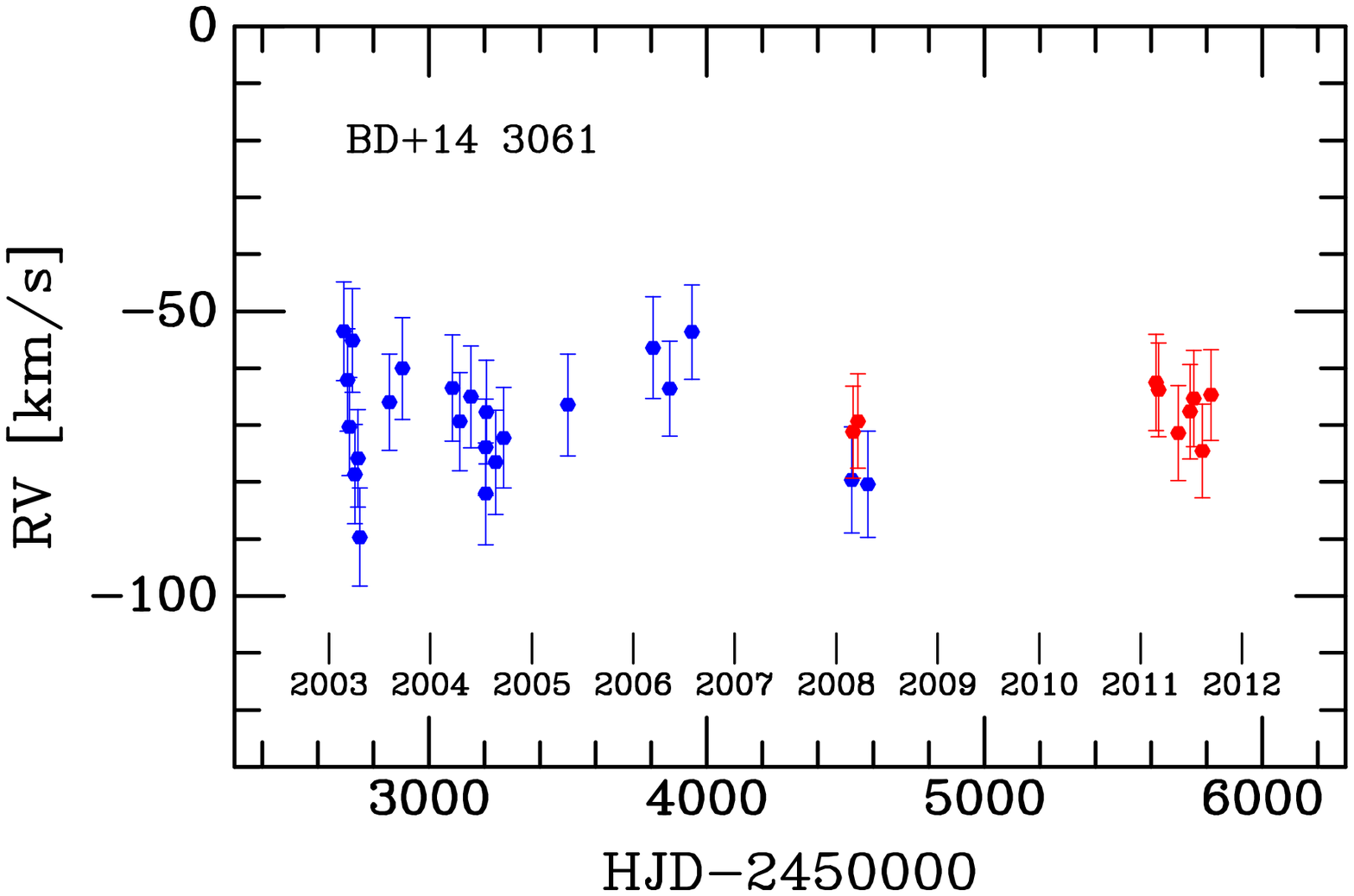}
\vskip0.2in
\includegraphics[width=0.47\textwidth]{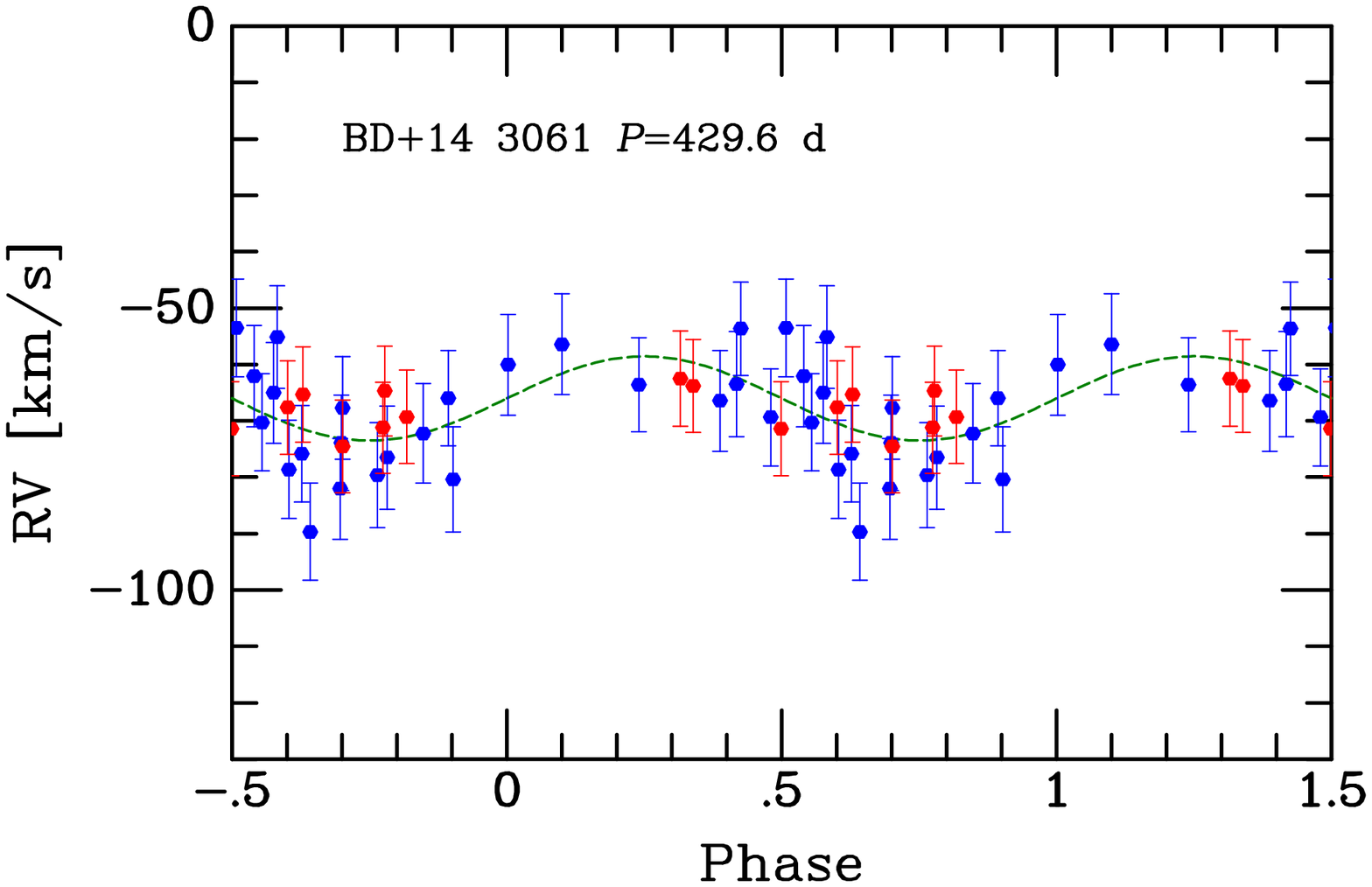}
\caption{
{\it Top panel:} Radial-velocity measurements of \bd\ from 2003 to 2011, from the SMARTS 1.5-m spectrograph. 
{\it Bottom panel:} The velocities phased with a period of 429.6~days. A sinusoidal fit to the phased velocities is overlain with a dashed green curve (see text for the parameters). In both plots, measurements made with the ``56/II''grating setup (2.2~\AA\ resolution) are in blue, and measurements with the ``47/IIb'' setup (1.6~\AA\ resolution) are in red.
\label{fig:rvs}
}
\end{figure}

I determined RVs from the spectra by cross-correlating with spectra of RV
standard stars taken on many of the same nights, using the procedures described
in detail in B16. The 1.5-m spectrograph was not highly optimized for precise RV
measurements; the typical uncertainty of a single measurement is about
$\pm$8--$9\,\kms$, with occasional larger outliers. Table~\ref{table:rvs} lists the results,
and the top panel in Figure~\ref{fig:rvs} plots the final velocities versus date. 

% The mean of the 32 measurements is $-68.49\pm1.5\,\kms$, with a standard
% deviation of $8.53\,\kms$. 

\begin{deluxetable}{lcc}
\tablewidth{0 pt}
\tablecaption{SMARTS 1.5-m Radial Velocities of BD+14$^\circ$3061
\label{table:rvs}}
\tablehead{
\colhead{HJD$-$2400000} &
\colhead{Radial Velocity} &
\colhead{Uncertainty} \\
\colhead{} &
\colhead{[$\kms$]} &
\colhead{[$\kms$]} 
}
\startdata
52692.90633  & $   -53.51 $ &         8.65 \\
52706.87252  & $   -62.03 $ &         8.99 \\
52712.89319  & $   -70.26 $ &         8.65 \\
52724.86802  & $   -55.10 $ &         9.09 \\
52733.88762  & $   -78.61 $ &         8.74 \\
52743.86193  & $   -75.86 $ &         8.57 \\
52750.85244  & $   -89.67 $ &         8.63 \\
52858.53605  & $   -65.95 $ &         8.44 \\
52905.46665  & $   -60.06 $ &         8.91 \\
53083.87481  & $   -63.48 $ &         9.32 \\
53110.82139  & $   -69.36 $ &         8.62 \\
53151.77894  & $   -65.02 $ &         8.95 \\
53203.63340  & $   -82.04 $ &         8.94 \\
53204.71839  & $   -73.89 $ &         8.46 \\
53205.63888  & $   -67.74 $ &         9.12 \\
53240.53756  & $   -76.53 $ &         9.11 \\
53268.48639  & $   -72.23 $ &         8.83 \\
53500.80060  & $   -66.42 $ &         8.93 \\
53806.86399  & $   -56.39 $ &         8.97 \\
53866.71382  & $   -63.57 $ &         8.35 \\
53946.63389  & $   -53.67 $ &         8.30 \\
54521.90710  & $   -79.61 $ &         9.33 \\
54580.84737  & $   -80.43 $ &         9.31 \\
54525.84424  & $   -71.21 $ &         8.07 \\
54544.80780  & $   -69.29 $ &         8.33 \\
55617.89389  & $   -62.51 $ &         8.45 \\
55627.86543  & $   -63.80 $ &         8.26 \\
55696.84913  & $   -71.39 $ &         8.31 \\
55740.69908  & $   -67.65 $ &         8.29 \\
55752.64857  & $   -65.35 $ &         8.47 \\
55783.54022  & $   -74.50 $ &         8.24 \\
55816.53672  & $   -64.70 $ &         7.96 \\
\enddata
\end{deluxetable}

The scatter in the RVs plotted in the top panel of Figure~\ref{fig:rvs} suggests that the RV
may be variable. A periodogram analysis yielded a most-likely period of about
430~days. I then applied a least-squares sinusoidal fit to the data, resulting
in a period of $P=429.6\pm6.4$~days, a semi-amplitude of $K_1=7.5\pm2.0\,\kms$,
and a center-of-mass RV of $\gamma=-66.0\pm1.6\,\kms$.  The $\gamma$-velocity is in
good agreement with the median RV of $-65.01\pm1.02\,\kms$ given by \Gaia\/ DR2.
The bottom panel in Figure~\ref{fig:rvs} shows the RVs phased with this period; superposed
is the sinusoidal fit. Given the relatively large uncertainties in the
individual velocities, the reality of the RV variations and this orbital period
must be regarded as tentative. Observations with higher precision are desirable
to confirm this result. Nevertheless the putative binary parameters are fairly similar to
those of the prototypical field yellow PAGB stars.

%Sine fit using /user/bond/klingsor1/lstsqs/sinewave_frontend.f: see 1.5m
%directory

For these binary parameters, the mass function is $f(m) \simeq 0.019\,M_\odot$. Assuming a mass for the PAGB star of $m_1\simeq0.53\,M_\odot$, I find a minimum mass for the unseen secondary of $m_2\simeq0.22\,M_\odot$. Thus the nature of the companion is fairly unconstrained; it could be a main-sequence star, or possibly a white dwarf.

\section{Summary}

In this paper I report my discovery, based on Str\"omgren photometry showing a very large Balmer discontinuity, that the metal-poor Galactic-halo field star \bd\ is a member of the rare class of yellow PAGB stars. From its \Gaia\/ DR2 parallax, I find a visual absolute magnitude of $-3.44\pm0.27$. This is close to the absolute brightness of a similar luminous star in the globular cluster M79, and supports the case that yellow PAGB stars may be useful Population~II standard candles.\footnote{ As the referee points out, if BD+14 is a binary with the period discussed here, its evolution should have been truncated by binary interactions and mass loss. Since the luminosities of PAGB stars are a function of core mass, such interruptions of their evolution will have imposed limits on their absolute magnitudes, which should vary from star to star. Nevertheless, the $M_V$ of BD+14 agrees very well with that of the yellow PAGB star in M79 (Figure~\ref{fig:hrd}), and with other such stars in GCs, as will be shown in forthcoming papers. Rigorous empirical tests will be required to establish yellow PAGB stars as standard candles for distance determination.}   The \Gaia\/ parallax and proper motion show that BD+14 has a very high space motion of $318\,\kms$, and moves in a retrograde Galactic orbit. The spectral-energy distribution of BD+14 out to $22\,\mu$m shows no evidence for circumstellar dust. Archival photometry reveals that BD+14 is a low-amplitude semi-regular pulsating variable, similar to the UU~Herculis class, with typical periods of 30--32~days. A radial-velocity study indicates a possible binary period of 429.6~days, which would make BD+14 similar to several other known long-period binaries among PAGB stars.

Useful future studies would include a high-dispersion abundance analysis, and confirmation of the binary period using more precise radial velocities. A more precise parallax from \Gaia\/ DR3 will strengthen its utility as a zero-point calibrator for yellow PAGB stars as standard candles.

\acknowledgments

I was motivated to publish this ancient discovery through discussions with Robin
Ciardullo, Brian Davis, and Michael Siegel. Penn State undergraduate Ben Hampton
carried out some of the radial-velocity data reduction. My research on
metal-deficient stars in the 1970s at Louisiana State University was partially
supported by National Science Foundation grant AST 78-25538.

I thank the STScI Director's Discretionary Research Fund for supporting 
participation in the SMARTS consortium, and Fred Walter for scheduling the
1.5-m  queue observations. I especially appreciate the excellent work of the
CTIO/SMARTS service observers who obtained the spectra during many long clear
Tololo nights:
% 1.5m data:
% additional from 1.3m
%   Claudio Aguilera,
%   Edgardo Cosgrove,
%   Juan Espinoza,
Sergio Gonz\'alez,
Manuel Hern\'andez, 
Rodrigo Hern\'andez,
%   Alberto Miranda,
Alberto Pasten,
%   Jacqueline Seron,
%   John Subasavage,
%   Joselino Vasquez,
and Jos\'e Vel\'asquez.
Daniel Maturana was my cheerful night assistant for the CTIO 0.9-m photometry.

This work has made use of data from the European Space Agency (ESA) mission {\it
Gaia\/} (\url{https://www.cosmos.esa.int/gaia}), processed by the {\it Gaia\/}
Data Processing and Analysis Consortium (DPAC,
\url{https://www.cosmos.esa.int/web/gaia/dpac/consortium}). Funding for the DPAC
has been provided by national institutions, in particular the institutions
participating in the {\it Gaia\/} Multilateral Agreement.

Based in part on observations made with the NASA {\it Galaxy Evolution Explorer}.
\GALEX\/ was operated for NASA by the California Institute of Technology under NASA
contract NAS5-98034.  
 
This research used data from the AAVSO Photometric All-Sky Survey (APASS),
funded by the Robert Martin Ayers Sciences Fund and NSF AST-1412587.

This publication makes use of data products from the Two Micron All Sky Survey,
which is a joint project of the University of Massachusetts and the Infrared
Processing and Analysis Center/California Institute of Technology, funded by the
National Aeronautics and Space Administration and the National Science
Foundation.

It also makes use of data products from the {\it Wide-field Infrared Survey
Explorer}, which is a joint project of the University of California, Los Angeles,
and the Jet Propulsion Laboratory/California Institute of Technology, funded by
the National Aeronautics and Space Administration.

\end{document}